# REDUCING THE COMPUTATIONAL REQUIREMENTS OF THE DIFFERENTIAL QUADRATURE METHOD

Wen Chen[a], Xinwei Wang[b] and Yongxi Yu[a]

[c] Mechanical Engineering Department, Shanghai Jiao Tong University, Shanghai 200030, P.R.China

[b] Department of Aircraft Engineering, Nanjing University of Aeronautics and Astronautics, Nanjing 210016, P. R. China.

The mail address for Wen Chen for proofs is P.O.Box 93B1, Shanghai Jiao Tong University, Shanghai 200030, P.R.China.

**Abstract**

This paper shows that the weighting coefficient matrices of the differential quadrature method (DQM) are centrosymmetric or skew-centrosymmetric if the grid spacings are symmetric irrespective of whether they are equal or unequal. A new skew centrosymmetric matrix is also discussed. The application of the properties of centrosymmetric and skew centrosymmetric matrix can reduce the computational effort of the DQM for calculations of the inverse, determinant, eigenvectors and eigenvalues by 75%. This computational advantage are also demonstrated via several numerical examples.

# 1. Introduction

The differential quadrature method (DQM) is a rather efficient numerical method for the rapid solution of linear and nonlinear partial differential equations involving one dimensions or multiple dimensions[1-9]. Compared with the standard methods such as the finite element and finite difference methods, the DQM requires less computer time and storage. The essence of the DQM is that a partial derivative of a function is approximated by a weighted linear sum of the function values at given discrete points. Its weighting coefficients do not relate to any special problem and only depend on the grid spacing. Thus, any partial differential equation can be easily reduced to a set of algebraic equations using these coefficients. The DQM coefficient matrices play a prominent role in the application of the DQM. Quan and Chang [10] pointed out that the skew centrosymmetric and centrosymmetric structures of the DQM weighting coefficient matrices for the 1st and 2nd order derivatives, and this properties can be used to decompose the DQM coefficient matrix for ordinary differential equations of systemic models into two smaller size matrices. But in that reference they did not use the centrosymmetric and skew centrosymmetric terminologies and thus can not be fully aware of the possible advantages in application of these properties. Chen and Yu [10] pointed out clearly the relation between the DQM weighting coefficient matrices and centrosymmetric or skew centrosymmetric matrices.

The present study is a further development of the above work. Firstly, we proves that the structure of the weighting coefficient matrices for any order derivative in the differential quadrature method are centrosymmetric or skew centrosymmetric if the grid spacing is symmetric with respect to the center point whether the grid spacings are equal or nonequal. It is known that, in the evaluation of the determinant, inverse and eigenvalue and eigenvectors, a centrosymmetric matrix can be factorized into two or three products, depending on whether the order N of the matrix is even or odd. By applying these properties, the multiplication complexity can be reduced by 75% and the efficiency of the DQM can be significantly increased, which is also demonstrated by solving the free vibration of beams and plates in this paper.

Secondly, the skew centrosymmetric matrix is defined and its properties are found to be similar to those of the centrosymmetric matrices. It is shown that this kind of matrix is related to the DQM weighting coefficient matrices for all odd order derivatives.

Next, the computational reduction by using the centrosymmetric properties in the DQM is extended to non-systemic problems.

Finally, based on the fact that DQM is a polynomial approach [8], a new truncation error estimation and its distribution in the entire variable domain are given in appendix 1.

## 2. Formulas for Direct Computing the Weighting Coefficients of the DQM

The differential quadrature method approximates the derivative of a function with respect to a variable at a given discrete point as a weighted linear sum of the function values at all discrete points. The general differential quadrature approximation at the ith discrete point is given by (Civan and Sliepcevich [4,5])

$$\frac{\partial^m f(x)}{\partial x^m}\bigg|_{x=x_i} = \sum_{j=1}^{N} w_{ij} f(x_j), \qquad i = 1, 2, \ldots, N. \tag{1}$$

where $x_j$ are the discrete points in the variable domain. $f(x_j)$ and $w_{ij}$ are the function values at these points and the weighting coefficients, respectively.

In order to determine the weighting coefficients $w_{ij}$, equation (1) must be exact for all polynomials of degree less than or equal to (N-1). To avoid the ill-conditioning Vandermonde matrix (see Bellman[12]) when N is relatively large, the Lagrange interpolation formula is used as a test function [6, 9, 10, 11], namely,

$$f_k(x) = \prod_{i \neq k}^{N} \frac{x - x_i}{x_k - x_i}. \tag{2}$$

Substituting Eq. (2) into Eq. (1) yields the following two formulae to compute directly the weighting coefficient of the 1st order derivative [6, 10], i.e.,

$$A_{ij} = \frac{1}{x_j - x_i} \prod_{\substack{k \neq i, j \\ i \neq j}}^{N} \frac{x_i - x_k}{x_j - x_k}, \qquad i = 1, 2, \ldots, N \text{ and } j = 1, 2, \ldots, N. \tag{3a}$$

and

$$A_{ii} = -\sum_{k \neq i}^{N} \frac{1}{x_i - x_k}, \qquad i = 1,2,\ldots,N. \tag{3b}$$

For higher order derivatives, the weighting coefficients can be generated by two recursion formulae [6]

$$w_{ij}^{(m+1)} = m(A_{ij}\, w_{ij}^{(m)} - \frac{w_{ij}^{(m-1)}}{x_i - x_j}), \qquad i \neq j \tag{4a}$$

and

$$w_{ii}^{(m+1)} = -\sum_{j \neq i}^{N} w_{ij}^{(m+1)}, \tag{4b}$$

where the superscript (m+1) denotes the order of the derivative.

## 3. The centrosymmetricity of weighting coefficient matrices in DQM

First, the definitions about centrosymmetric and skew centrosymmetric matrices are given

Definition 1: A N×N matrix Q=[$q_{ij}$] is centrosymmetric if

$$q_{ij} = -q_{N+1-i, N+1-j}, \quad i, j=1, 2, \ldots, N. \tag{5}$$

Then, Q can be characterized by

$$JQJ = Q, \tag{6}$$

where J denotes the contra-identity matrix given by

$$J = \begin{bmatrix} 0 & \cdots & 0 & 1 \\ 0 & \cdots & 1 & 0 \\ \vdots & \ddots & \vdots & \vdots \\ 1 & \cdots & 0 & 0 \end{bmatrix}, \tag{7}$$

which has unit elements along the secondary diagonal and zeros elsewhere, noting that $J^T=J$ and $J^2=I$, the unit matrix, The effect of premultiplying any matrix by J reverses the order of its rows, and postmultipling reverses the order of its columns.

Definition 2: A new, skew centrosymmetric matrix $R=[r_{ij}]_{N \times N}$ with the following property, is introduced:

$$r_{ij} = -r_{N+1-i, N+1-j}, \quad i,j=1,2,\ldots,N. \tag{8}$$

and

$$R = -JRJ. \tag{9}$$

To our knowledge, this is a new form of matrix, which has not been reported previously. In the following, the structure of weighting coefficient matrices of the DQM is analyzed. By applying Eq. (3a) and (3b), it can be deduced that

$$A_{ij} = \frac{1}{x_j - x_i} \prod_{\substack{k \neq i, j \\ i \neq j}}^{N} \frac{x_i - x_k}{x_j - x_k}, \quad i, j = 1, 2, \ldots, N \tag{10a}$$

and

$$A_{ii} = \sum_{k \neq i}^{N} \frac{1}{(1-x_i)-(1-x_k)}, \quad i = 1, 2, \ldots, N. \tag{10b}$$

The grid spacings often used in DQM are symmetric such as equally spaced grids and the roots of the shifted Legendre or the shifted Chebyshev polynomials. For symmetric grid spacing, over a domain $0 \leq x \leq 1$, it is true that:

$$x_{N+1-i} = 1 - x_i \tag{11}$$

where N is the number of grid points. Thus, substituting Eq. (11) into (10a) and (10b) yield

$$A_{ij} = \frac{1}{x_{N+1-j} - x_{N+1-i}} \prod_{\substack{k \neq N+1-i, N+1-j \\ i \neq j}}^{N} \frac{x_{N+1-i} - x_k}{x_{N+1-j} - x_k}, \quad i, j = 1, 2, \ldots, N \tag{12a}$$

and

$$A_{ii} = \sum_{k \neq N+1-i}^{N} \frac{1}{x_{N+1-i} - x_k}, \quad i = 1, 2, \ldots, N. \tag{12b}$$

Thus,

$$A_{ij} = -A_{N+1-i, N+1-j}. \tag{13}$$

According to the definition given by Eq. (8), the weighting coefficient matrix for the 1st order derivative can be concluded to be skew centrosymmetric if the grid spacing is symmetric. Next, consider the coefficient matrix for the high order derivative. Using Eqs. (13) and the recursion formulas (4a) and (4b) and after some manipulations, it can be shown that

$$w_{ij}^{(m+1)} = (-1)^{(m+1)} w_{N+1-i, N+1-j}^{(m+1)}. \tag{14}$$

where the $w_{ij}^{(m+1)}$ denotes the DQM coefficient for the (m+1)th order derivative. Therefore, the DQM weighting coefficient matrices are skew centrosymmetric for odd order derivatives and centrosymmetric for even order derivatives when the grid spacing is symmetric.

## 4. Some Properties of the Centrosymmetric and Skew Centrosymmetric Matrices

### 4.1 Centrosymmetric Matrices

For completeness, the interesting structure of the centrosymmetric matrix of even degree [13,14] is described briefly. The corresponding similar properties exist for odd order centrosymmetric matrix. Here $C_{N \times N}$ denotes the set of N×N centrosymmetric matrices.

Lemma 1. If $Q_1, Q_2 \in C_{N \times N}$ and $Q_1^{-1}$ exists, then

(1) $Q = Q_1 Q_2 \in C_{N \times N}$

(2) $Q = Q_1 + Q_2 \in C_{N \times N}$ (15)

(3) $Q_1^{-1} \in C_{N \times N}$

Lemma 2. If $Q \in C_{N \times N}$ and N=2M, Q can be written as

$$Q = \begin{bmatrix} A & JCJ \\ C & JAJ \end{bmatrix},$$ (16)

where A and C are arbitrary M×M matrices. The determinant of matrix Q can be evaluated by

$$|Q| = |A + JC||A - JC|$$ (17)

and the inverse of the matrix is given as

$$Q = \begin{bmatrix} P & JRJ \\ R & JPJ \end{bmatrix}, \tag{18}$$

where $2P=(A+JC)^{-1}+(A-JC)^{-1}$ and $2R=(A+JC)^{-1}-(A-JC)^{-1}$.

Applying Eqs. (17) and (18), the calculation effort of the inverse and determinant of a centrosymmetric matrix can be reduced by 75%. Note that premultiplying or postmultiplying any matrix by J only moves the elements of the matrix and requires a very little computational time.

Lemma 3. If $Q \in C_{N \times N}$ and N=2M, then

$$Q = \begin{bmatrix} A & JCJ \\ C & JAJ \end{bmatrix} \quad \text{and} \quad \begin{bmatrix} A-JC & 0 \\ 0 & A+JC \end{bmatrix} \tag{19}$$

are orthogonally similar. Thus, the evaluation of the eigenvectors and eigenvalues of Q is equivalent to the that of two M×M matrices A-JC and A+JC.

Lemma 4. The eigenvector of centrosymmetric matrix Q in Lemma 3 is the sum of a symmetric vector and a skew-symmetric vector. The N/2 skew symmetric orthonormal eigenvectors $U_i$ and the corresponding eigenvalues $\lambda_i$ can be determined by

$$(A - JC)v_i = \lambda v_i \tag{20a}$$

and

$$U_i = \frac{1}{\sqrt{2}}\left[y_i^T \quad -(Jv_i)^T\right].\tag{20b}$$

The N/2 symmetric eigenvectors $W_i$ and the corresponding eigenvalues $P_i$ are found by solving the following equation:

$$(A+JC)y_i = P y_i \tag{21a}$$

and

$$W_i = \frac{1}{\sqrt{2}}\left[y_i^T \quad (Jy_i)^T\right] \tag{21b}$$

**4.2 Skew Centrosymmetric Matrices**

Next, the structure of skew centrosymmetric matrix is discussed. The proofs for these properties are similar to those for centrosymmetric matrices [13, 14], so they are omitted here for brevity. We herein analysis only the even order skew centrosymmetric matrix, but it should be pointed out that there are similar properties for odd order skew centrosymmetric matrix. In the following, $NC_{N\times N}$ is the set of N×N skew centrosymmetric matrices.

Lemma 5. If $Q_1$, $Q_2 \in NC_{N\times N}$ and $Q_1^{-1}$ exists, then

1) $Q=Q_1Q_2 \in C_{N\times N}$

(2) $Q=Q_1+Q_2 \in NC_{N\times N}$ (22)

(3) $Q_1^{-1} \in NC_{N\times N}$

Lemma 6. If $Q \in NC_{N \times N}$ and $N=2M$, then $Q$ can be stated as

$$Q = \begin{bmatrix} A & -JCJ \\ C & -JAJ \end{bmatrix} \tag{23}$$

where A and C are arbitrary $M \times M$ matrices. The determinant of matrix Q can be expressed as

$$|Q| = |A+JC||A-JC| \tag{24}$$

and the inversion of the matrix is given as

$$Q = \begin{bmatrix} P & -JRJ \\ R & -JPJ \end{bmatrix} \tag{25}$$

where $2P=(A+JC)^{-1}+(A-JC)^{-1}$ and $2R=(A+JC)^{-1} - (A-JC)^{-1}$.

Lemma 7. If $Q \in NC_{N \times N}$ and $N=2M$, then

$$Q = \begin{bmatrix} A & -JCJ \\ C & -JAJ \end{bmatrix} \quad and \quad \begin{bmatrix} 0 & A-JC \\ A+JC & 0 \end{bmatrix} \tag{26}$$

are orthogonally similar. Thus, the evaluation of the eigenvectors and eigenvalues of Q is equivalent to the that of an $M \times M$ matrices.

## 5. Application

In this section, several examples are solved by the DQM and the aforementioned properties of the DQM coefficient matrix are applied to reduce the computational effort. There are several approaches available in applying the DQM for high order boundary value problems.

In this paper, according to Wang and Bert [8], the weighting coefficient matrices are modified in terms of the specific boundary conditions. It is obvious that such coefficient matrices are also centrosymmetric when the boundary conditions are symmetric. In the applications, equally spaced points and the roots of the shifted Chebyshev polynomials given below are adopted as the grid points.

$$x_i = \cos\frac{(2i-1)\pi}{2N}, \quad i = 1, 2, \ldots, N. \tag{27}$$

For the later, the following formulae can be derived.

$$A_{ii} = \frac{x_i}{(1-x_i^2)}, \quad i = 1, 2, \ldots, N. \tag{28a}$$

and

$$A_{ij} = \frac{(-1)^{(i-j)}}{(x_i - x_j)}\sqrt{\frac{1-x_j^2}{1-x_i^2}}, \quad i \neq j, \tag{28b}$$

where $A_{ij}$'s represent the weighting coefficients of the 1st order derivative. The weighting coefficients for higher order derivatives can be easily computed from Eqs (4a) and (4b).

Example 1: Flexural Vibration of a Simply Supported Beam

The governing differential equation for this example can be expressed as:

$$W^{iv} = \varpi^2 W(x), \tag{29a}$$

where the nondimensionalized frequency is $\varpi^2 = \rho A_0 L^4 \omega^2/EI$. $A_0$, L and $\rho$ are the constant cross-sectional area, the length of the beam, the density and I the constant area moment of inertia about the neutral axis, respectively. The boundary conditions at the simply supported ends are given by:

$w(0) = 0, \quad w''(0) = 0$, (29b)

$w(1) = 0, \quad w''(1) = 0$. (29c)

In terms of differential quadrature, Eq. (29a) is expressed as:

$$\sum_{j=2}^{N-1} \overline{D}_{ij} W_j = \varpi^2 W_i, \quad i = 2, \ldots, (N-1).$$ (29d)

Note that the boundary conditions have been applied in the formulation of the weighting coefficient matrix $\overline{D}_{ij}$, which is a centrosymmetric matrix as shown in the appendix 2. Eq. (29d) can be written in the matrix form as:

$$\begin{bmatrix} \overline{D}_{22} & \overline{D}_{23} & \cdots & \overline{D}_{2,N-1} \\ \overline{D}_{32} & \overline{D}_{33} & \cdots & \overline{D}_{3,N-1} \\ \vdots & \vdots & \vdots & \vdots \\ \overline{D}_{N-1,2} & \overline{D}_{N-1,3} & \cdots & \overline{D}_{N-1,N-1} \end{bmatrix} \begin{bmatrix} w_2 \\ w_3 \\ \vdots \\ w_{N-1} \end{bmatrix} = \varpi^2 \begin{bmatrix} w_2 \\ w_3 \\ \vdots \\ w_{N-1} \end{bmatrix}.$$ (29e)

According to Lemma 2, we have

$$\left\{ \overline{D}_{ij} \right\}_{(N-2)\times(N-2)} = \begin{bmatrix} P & JRJ \\ R & JPJ \end{bmatrix}.$$ (29f)

In the present study, eight equally spaced grid points are used. Therefore, the order of the matrices P and R is 3. In addition, it is known that the eigenvectors corresponding to the fundamental frequency of a simply supported beam is symmetric. Thus, according to lemma 4, the fundamental frequency and the corresponding eigenvector can be computed by:

$$(P+JR)u_i = \varpi^2 u_i. \tag{29g}$$

Similarly, the 2nd order frequency with skew-symmetric eigenvector can be calculated by:

$$(P - JR)v_i = \varpi^2 v_i \tag{29h}$$

The fundamental frequency obtained by the present method is 9.8683, comparing well with the exact solution of 9.8696. The relative error is -0.006%. The 2nd order frequency is 39.2411, and the exact solution 39.4784, hence the relative error is 0.6%. If a nonuniform grid formed from the Chebyshev polynomials are used, the errors are further reduced to 0.000% for the fundamental frequency and to -0.17% for the 2nd frequency. In contrast with Wang and Bert [8], the present DQM reduces the computational effort by 87.5% because of the use of the centrosymmetric property of weighting coefficients matrix.

Example 2: Transverse vibration of a thin rectangular plate simply supported on all four edges.

Applying the DQM, the governing equation [7] for this case is given by:

$$\sum_{k=2}^{N-1} \bar{D}_{ik} W_{kj} + (2\alpha^2) \sum_{m=2}^{N-1} \bar{B}_{jm} \sum_{k=2}^{N-1} \bar{B}_{ik} W_{km} + (\alpha^4) \sum_{k=2}^{N-1} \bar{D}_{jk} W_{ik} = \varpi^2 W_{ij}, \tag{30}$$

where $\varpi^2 = \rho h a^4 \omega^2/D$, and $\alpha = a/b = 1.5$, and $\bar{D}$ and $\bar{B}$ are the modified weighting coefficient matrices for the fourth and second derivatives, respectively.

The DQM with a rather small number of grid points can produce very good results. In this case, 8×8 equally spaced grids are used. The DQM results for the fundamental of $\varpi=32.0721$ and for the 2nd frequency of $\varpi=61.4449$ agree well with the exact solutions of 32.0762 and 61.6850. The relative error is -0.01% and -0.39%. If the roots of the shifted Chebyshev polynomials are used as the grid points, the DQM solutions for 1st and 2nd frequency are calculated as 32.0761 and 61.6159, and the errors are further reduced to -0.0003% and 0.11%. For the same problem, Wang and Bert [8] need compute a $(N-2)^2 \times (N-2)^2$ order matrix eigenvalue problem. In contrast, the present approach only requires solving a $(N-2)^2/2 \times (N-2)^2/2$ matrix due to the use of the DQM centrosymmetric properties. Therefore, the computational effort are only 12.5% of that of Wang and Bert [8].

**Example 3. The transverse vibration of a thin, isotropic, skew plate vibration**

This problem have been handled by the DQ method in reference [15]. But since its governing and boundary condition equations contain cross derivative term, and thus its DQ formulation exists the centrosymmetric and skew centrosymmetric matrices simultaneously, we analyze this problem again to demonstrate the computational reduction by the centrosymmetric properties for this sort of problem. The governing equation for this case is given by

$$w_{,\xi\xi\xi\xi} - (4\cos\theta) w_{,\xi\xi\xi\eta} + 2(1 + 2\cos^2\theta) w_{,\xi\xi\eta\eta} - (4\cos\theta) w_{,\xi\eta\eta\eta} + w_{,\eta\eta\eta\eta} = (\rho h \omega^2 / D) \sin^4\theta \quad (31a)$$

where θ=skew angle, ρ density, ω circular natural frequency and D flexural rigidity. The simply supported boundary conditions are

$$w = 0, \quad w_{,\xi\xi} - 2w_{,\xi\eta}\cos\theta = 0, \quad at \quad \xi = 0, a$$
$$w = 0, \quad w_{,\eta\eta} - 2w_{,\xi\eta}\cos\theta = 0, \quad at \quad \eta = 0, b \quad (31b)$$

The clamped boundary conditions are

$$w = 0, \quad w_{,\xi} = 0, \quad at \quad \xi = 0, a$$
$$w = 0, \quad w_{,\eta} = 0, \quad at \quad \eta = 0, b \quad (31c)$$

let $x=(2\zeta-a)/a$, $y=(2\eta-b)/b$. The matrix formulations for derivatives of a two-variable function are presented herein, i.e.

$$\frac{\partial^4 w}{\partial x^4} = \underline{D}_x w, \quad \frac{\partial^4 w}{\partial x^2 \partial y^2} = \underline{B}_x w \underline{B}_y^T, \quad \frac{\partial^4 w}{\partial y^4} = w \underline{D}_y^T,$$
$$\frac{\partial^4 w}{\partial x^3 \partial y} = \underline{C}_x w \underline{A}_y^T, \quad \frac{\partial^4 w}{\partial x \partial y^3} = \underline{A}_x w \underline{C}_y^T \quad (31d)$$

Note that W is (n-2)×(n-2) matrix. The DQM formulation for the governing equation (31a) can be stated as

$$\underline{D}_x W - (4\beta\cos\theta)\underline{C}_x w \underline{A}_y^T + 2\beta^2(1 + 2\cos^2\theta)\underline{B}_x w \underline{B}_y^T -$$
$$(4\beta\cos\theta)\underline{A}_x w \underline{C}_y^T + \beta^4 w \underline{D}_y^T = (\omega_l^2/16)w \quad (31f)$$

Applying the kronecker product of matrix to the above equation, we have

$$\{[I_{n-2}] \otimes [\underline{D}_x] - (4\beta\cos\theta)[\underline{A}_y] \otimes [\underline{C}_x] + 2\beta^2(1 + 2\cos^2\theta)[\underline{B}_y] \otimes [\underline{B}_x] -$$
$$(4\beta\cos\theta)[\underline{C}_y] \otimes [\underline{A}_x] + \beta^4[\underline{D}_y] \otimes [I_{n-2}]\}vec\{W\} = (\omega_l^2/16)vec\{W\} \quad (31g)$$

where ⊗ denotes the kronecker product in matrix computation (the DQ formulation for this case in Eq. (10) of the reference [15] was incorrectly expressed). $[I_{n-2}]$ represents a unite matrix, whose n is the number of grid points. vec(w) is the vector-function of a rectangular matrix formed by stacking the columns of matrix into one long vector. $\varpi^2 = (\rho h\, a^4\, \omega^2 / D) \sin^4 \theta$. β and D are aspect ratio and flexural rigidity, respectively. Weighting coefficient matrices $\overline{A}, \overline{C}, \overline{D}$ and $\overline{B}$ are modified by the boundary conditions. For clamped boundary conditions, the $\overline{D}$ and $\overline{B}$ are the same as the rectangular plate. But the problem is more complex for simply supported boundary conditions because it contain cross derivative term $w_{,xy}$. Using the DQ method, we have $w_{,xy} = A_x\, w\, A_y^T = (A_y \otimes A_x)\, vec(w)$. It is easily proved that $A_x \otimes A_y$ is a centrosymmetric matrix, since $A_x$ and $A_y$ are the skew centrosymmetric matrices in using symmetric distributed grid points as in reference [15]. Therefore, $\overline{D}$ and $\overline{B}$ will be centrosymmetric matrices and $\overline{A}, \overline{C}$ the skew centrosymmetric matrix. Obviously, the resulting matrix for the governing equation (31g) is a centrosymmetric matrix. Thus, we can apply the computational reduction of centrosymmetric matrix for this case similar to example 2. Namely, the computational effort is reduced by 87.5%.

**Example 4. The convection-diffusion problem**

Civan and Sliepcevich [16] computed the convection-diffusion problems by the differential quadrature method. We analyze this example again to show that the computational reduction in centrosymmetric matrix can be used for those problems with non-symmetric boundary conditions. We consider the steady-state case. In terms of the DQ method, the governing equation [16] can be expressed as

$$-\frac{\phi_{ij}}{4a} + \alpha \sum_{k=1}^{N^x} b_{ik}{}^x \phi_{kj} + \beta \sum_{k=1}^{N^y} b_{jk}{}^y \phi_{jk} = 0, \qquad (32a)$$

where $\alpha$ and $\beta$ are constant quantities. $b_{ik}$'s are the DQ weighting coefficient for the second derivative and have been not modified by boundary conditions, which is different from example 1-3. The boundary conditions are approximated by the DQ method as

$\phi_{ij}$=prescribed value, j=1,2, ..., $N^y$ (32b)

$\phi_{iN^y} = 0$, j=1,2, ..., $N^x$ (32c)

$$\frac{\partial \phi_{ij}}{\partial y} = \sum_{k=1}^{Ny} a_{jk}{}^y \phi_{ik} = 0, \text{ at j=1.} \qquad (32d)$$

$\phi_{N^x j} = 0$, j=1,2, ..., $N^x$ (32c)

Since the boundary conditions in this problem are not systemic, we need handle with Eq. (32a) and Eq. (32b-c), respectively. Obviously, the resulting coefficient matrix in Eq. (32a) is a centrosymmetric matrix irrespective of whether the boundary conditions are symmetric or non-symmetric. Solving the Eq. (32a) by using the centrosymmetric properties, the function values at inner grid points are expressed by the side lateral function values using a matrix formula. Substituting this formula into the boundary condition equations (32b-c), we can obtain the algebraic equations only containing the side lateral function values. The order of this equation is much smaller than that of Eq. (32a). The computational effort for solving it will be little. The main computational effort in this procedure is solving Eq. (32a), while the computational reduction by using the centrosymmetric matrix can be achieved in this step. So the present computational effort is nearly only 25% as that in reference [16].

For the deflection and vibrational analysis of plates and beams with non-systemic boundary conditions, the present reduction approach is also applicable, but the approach presented by Wang and Bert [8] in applying the DQM to structural problems can be not used for the cases. Another two different approaches proposed separately by Bert et. al. [7, 17] and by Chen et. al [18] should be used, in which the governing and boundary condition equations can be handled with separately.

## 6. Conclusion

This paper proves generally the centrosymmetricity of the weighting coefficient matrices in the DQM when the grid spacing is symmetric. Also a new type of matrix, called as the skew centrosymmetric matrix, is defined and its main properties are discussed. The structure of such matrices allows for factorization of the determinant and the characteristic equations into two, smaller and nearly equal size matrices. Therefore, the computational complexity can be reduced by 75%. Furthermore, it is shown that these matrices possess either symmetric or skew-symmetric eigenvectors. In many problems, the need of symmetric or skew-symmetric eigenvectors can be applied as a constraint. Therefore, the computational effort can be reduced further, as demonstrated by the foregoing examples.

The present work emphasizes the importance of the analysis of the weighting coefficient matrices in the DQM and makes the DQM even more attractive for the practical engineering analysis.

## Appendix 1

**The DQM Truncation Error**

If a function can be approximated by a Lagrangian polynomial [1, 9], then

$$f(x) = \psi_j(x) f(x_j) + R(x), \quad j = 1, 2, \ldots, N. \quad (A1)$$

where summation convection is applied on the repeated index j, $\psi_j(x)$ are the Lagrangian polynomials, and $R(x)$ is the truncation error. Hence,

$$f'(x_i) = \psi'_j(x_i) f(x_j) + R'(x_i)$$
$$= A_{ij} f(x_j) + R'(x_i), \quad I, j=1,2,\ldots,N. \quad (A2)$$

where $A_{ij}$ are the DQM weighting coefficient of the 1st order derivative. $x_i$ and $x_j$ are the discrete interpolation points, $R'(x_i)$ is the truncation error of the DQM to approximate 1st order derivative. The truncation error for higher order derivatives can be obtained in a similar way. Note that

$$R(x) = \frac{f^{(N)}(\zeta)W(x)}{N!}, \tag{A3}$$

where $W(x) = \prod_{i=1}^{N}(x - x_i)$. Thus,

$$R(x_i) = \frac{f^{(N)}(\zeta)W'(x_i)}{N!}, \quad i=1,2,...,N. \tag{A4}$$

It should be pointed out that the truncation error formula (A4) is different from that of Bellman [1]. Considering equally spaced grid points and assuming $|f^{(N)}(\zeta)| \leq K$. Then,

$$|R'(x_i)| \leq \frac{K}{N(N-1)^{N-1}} \quad \text{and} \quad |R'(x_N)| \leq \frac{K}{N(N-1)^{N-1}} \tag{A5}$$

for both ends and

$$|R'(x_{N/2})| \leq \frac{(N/2-1)!(N/2)!K}{N!(N-1)^{N-1}} \tag{A6}$$

for the center points if N is an even number.

Comparing the truncation error formulae (A5) for the ends with that for the center point (A6), it may be concluded that largest truncation error could possibly occur in the vicinity of the ends if equally spaced grid points are used in the DQM. The two examples as accuracy tests (Eqs. (4) and (5) of ref. 4) can be as the numerical computation demonstration for the present error analysis (see fig. 1 and fig. 2 of ref. 4). Also a similar error distribution is expected for unequally spaced grids such as defined by the roots of the shifted Chebyshev polynomials.

## Appendix 2

**Application of the boundary conditions in the formulation of the weighting coefficients**

Reference [8[ provided a new approach in applying the boundary conditions in the DQM, which applied the boundary conditions in formulation of the weighting coefficients. For a simply supported beam, reference [8] gave a detailed description. When the symmetric grid spacing is used, obviously, the modified 1st order derivative weighting coefficient matrix $[\overline{A}]$ (Eq. (3) of ref. 8) is a skew centrosymmetric matrix like [A] (Eq. (2) of ref. 8). The modified weighting coefficient matrix for the 2nd order derivative is given by (Eq. (4) of ref. 8)

$$[\overline{B}] = [A][\overline{A}]. \tag{B1}$$

According to Lemma 5, $[\overline{B}]$ is a centrosymmetric matrix. The modified weighting coefficient matrix for the 4th order derivative $[\overline{D}]$ is decided by (Eq. (6) of ref. 8)

$$[\overline{D}] = [\overline{B}][\overline{B}]. \tag{B2}$$

According to Lemma 1, $[\overline{D}]$ is also a centrosymmetric matrix.